\documentclass[prl,reprint,twocolumn,superscriptaddress,amsmath,amssymb,floatfix]{revtex4-2}

\usepackage[utf8]{inputenc}
\usepackage{graphicx}
\usepackage[usenames,dvipsnames]{xcolor}
\usepackage{epstopdf}
\usepackage{amsmath,amsthm,amssymb}
\usepackage{latexsym}
\usepackage{bm}
\usepackage{dcolumn}
\usepackage{braket}
\usepackage[normalem]{ulem}
\usepackage{times}
\usepackage{hyperref}
\usepackage[section]{placeins}

\newcommand{\yms}{YbMnSb$_{2}$}
\newcommand{\ymb}{YbMnBi$_{2}$}
\newcommand{\cmb}{CaMnBi$_{2}$}
\newcommand{\smb}{SrMnBi$_{2}$}

\newcommand{\bq}{\mbox{\boldmath$q$}}
\newcommand{\bQ}{\mbox{\boldmath$Q$}}

\begin{document}
\bibliographystyle{apsrev4-2}

\title{Coupling of magnetism and Dirac fermions in \yms\ }

\author{Xiao Hu}
%\email{xhu@bnl.gov}
\affiliation{Condensed Matter Physics and Materials Science Division, Brookhaven National Laboratory, Upton, NY 11973, USA}

\author{Aashish Sapkota}
%\email{sapkota@bnl.gov}
%\email{asapkota@ameslab.gov}
\affiliation{Condensed Matter Physics and Materials Science Division, Brookhaven National Laboratory, Upton, NY 11973, USA}
\affiliation{Ames National Laboratory, U.S. DOE, Department of Physics and Astronomy, Iowa State University, Ames, Iowa 50011, USA}
% 2Department of Physics and Astronomy, Iowa State University, Ames, Iowa 50011, USA

\author{Zhixiang Hu}
%\email{zhixiang@bnl.gov}
\affiliation{Condensed Matter Physics and Materials Science Division, Brookhaven National Laboratory, Upton, NY 11973, USA}
\affiliation{Department of Material Science and Chemical Engineering, Stony Brook University, Stony Brook, NY, 11790, USA}

\author{Andrei T. Savici}
%\email{saviciat@ornl.gov}
\affiliation{Neutron Scattering Division, Oak Ridge National Laboratory, Oak Ridge, TN 37831, USA}

\author{Alexander I. Kolesnikov}
%\email{kolesnikovai@ornl.gov}
\affiliation{Neutron Scattering Division, Oak Ridge National Laboratory, Oak Ridge, TN 37831, USA}

\author{John M. Tranquada}
%\email{jtran@bnl.gov}
\affiliation{Condensed Matter Physics and Materials Science Division, Brookhaven National Laboratory, Upton, NY 11973, USA}

\author{Cedomir Petrovic}
%\email{petrovic@bnl.gov}
\affiliation{Condensed Matter Physics and Materials Science Division, Brookhaven National Laboratory, Upton, NY 11973, USA}
\affiliation{Department of Material Science and Chemical Engineering, Stony Brook University, Stony Brook, NY, 11790, USA}

\author{Igor A. Zaliznyak}
\email{zaliznyak@bnl.gov}
\affiliation{Condensed Matter Physics and Materials Science Division, Brookhaven National Laboratory, Upton, NY 11973, USA}

\begin{abstract}
{We report inelastic neutron scattering measurements of magnetic excitations in \yms, a low-carrier-density Dirac semimetal in which the antiferromagnetic Mn layers are interleaved with Sb layers that host Dirac fermions. We observe a measurable broadening of spin waves, which is consistent with substantial spin-fermion coupling. The spin wave damping, $\gamma$, in \yms\ is roughly twice larger compared to that in a sister material, \ymb, where an indication of a small damping consistent with theoretical analysis %predicting subtle manifestation
of the spin-fermion coupling was reported. The inter-plane interaction between the Mn layers in \yms\ is also much stronger, suggesting that the interaction mechanism is rooted in the same spin-fermion coupling. Our results establish the systematics of spin-fermion interactions in layered magnetic Dirac materials.
% and is consistent with $\gamma \sim J^2$, $J_c^2~\gamma^2$ scaling that could be expected in such case
}

%\bigskip
%\newpage
%\emph{One sentence summary:}
%Our neutron scattering study of magnetic excitations in Dirac semimetal, \YMS, demonstrates ... .

\end{abstract}

\date{\today}

\maketitle
\newpage

\noindent\emph{Introduction.}
Dirac semimetals remain at the forefront of research on topological materials because of the fascinating quantum electronic phenomena they exhibit and of their potential technological applications \textcolor{blue}{\cite{Katmis_Nature2016,Masuda_2016,Khodas_2009,Wehling_2014,Kefeng_2011,Aifeng_2016}}. In these materials, the characteristic linear electronic dispersion leads to novel behaviors such as spin-polarized transport \textcolor{blue}{\cite{Khodas_2009}}, suppression of back-scattering due to spin-momentum locking \textcolor{blue}{\cite{Konig_2007,Hsieh_2009,Hor_2010}}, the chiral anomaly \textcolor{blue}{\cite{Ezawa_2017,Lv_2017,Aji2012}}, impurity-induced resonant states, and the anomalous quantum Hall effect \textcolor{blue}{\cite{Wehling_2014,Kefeng_2011,Aifeng_2016,Zhang_2005,Hasan_2010}}.

Among different types of Dirac semimetals, the family of 112 ternary pnictogens with the general formula $A/R$Mn$X_2$ ($A=$ Ca, Sr; $R=$ Yb, Eu; $X=$ Bi, Sb) have attracted particular attention due to the combination of highly anisotropic Dirac dispersion in quasi-2D square nets of $X$ atoms and strongly correlated magnetism of Mn \textcolor{blue}{\cite{Park_2011,Wang_2012,Wang_2016,Kefeng_2011,May2014,Chinotti_2016,Lee_2013,Liu2016,Chaudhuri_2017,Kealhofer_etal_PRB2018,Wang_PRM2018}}. These materials feature a common layered structure in which the $X$ layers hosting itinerant Dirac charge carriers are separated by strongly-correlated insulating Mn-$X$ layers. Both the inter-layer charge transport and the magnetic correlations between the Mn layers require that Dirac carriers are coupled to strongly-correlated Mn electrons. Therefore, these materials have become a fertile playground for investigating the interaction of the conduction Dirac electrons with the local-spin magnetic Mn-$X$ sublattice, i.e. spin-Dirac fermion coupling \textcolor{blue}{\cite{Kefeng_2011,Wang_2016,Park_2011,Zhang_2016,Guo_2014}}.

Previous inelastic neutron scattering (INS) measurements on (Sr, Ca)MnBi$_2$ reported no indication of such coupling because anomalous broadening of magnetic excitations found in itinerant magnets was not observed \textcolor{blue}{\cite{Guo_2014,Rahn_2017}}. Yet, the out-of-plane antiferromagnetism in SrMnBi$_2$ and ferromagnetism in CaMnBi$_2$ \textcolor{blue}{\cite{Guo_2014}} clearly indicate the presence of inter-layer interaction between magnetic Mn$^{2+}$ ions, which inevitably involves Dirac electrons in the interweaving Bi square nets. A detailed analysis of high-resolution INS measurements of magnetic excitations in \ymb\ led us to discover a signature of spin-Dirac fermion coupling in this material \textcolor{blue}{\cite{Sapkota2020}}. We found a small but distinct broadening of spin wave dispersion, both for the in-plane and the out-of-plane directions. For $T<T_N$, the broadening is weakly dependent on temperature and is nearly \bQ-independent. By comparing the observed spin wave damping with theoretical model of Dirac fermions coupled to spin waves, we found a very substantial spin-fermion coupling parameter, $g\approx 1.0$~eV$^{3/2}$ $\text\AA$, implicated in the theoretical description.

In order to establish the systematics of spin-fermion coupling in the 112 family of Dirac semimetals and further elucidate its properties, we carried out INS measurements on a sister material, \yms, where heavier Bi is substituted with the lighter Sb, thus reducing the spin-orbit coupling (SOC) and potentially softening Dirac dispersion. \yms\ crystalizes in the same $P4/nmm$ space group as \ymb, but weaker SOC is more favorable for stronger coupling of the massless Dirac fermions to magnons \textcolor{blue}{\cite{Liu2017,Liu2016}}. From the analysis of well-defined magnetic excitations observed in our experiments, we extract a damping parameter consistent with appreciable broadening of spin waves and substantial spin-fermion coupling. The spin wave damping and the inter-layer interaction in \yms\ are significantly stronger than those in \ymb. We note that for our measurements at low temperature of $\approx 5.5$~K,
%, which is much lower than the N\'eel temperature, $T_{\rm N}\approx$ 345 K \textcolor{blue}{\cite{Wang_PRM2018}},
damping induced by spin-phonon coupling is greatly suppressed and thus our observations corroborate the idea that it originates from coupling to Dirac fermions.

\noindent\emph{Experimental Details.}
Single crystals of \yms\ were grown from Sb flux using the method described in \textcolor{blue}{\cite{Wang_PRM2018}}. \yms\ orders antiferromagnetically below $T_\mathrm{N} \approx 345$~K, with an ordered moment of $3.48\mu_\mathrm{B}$ at 2~K \textcolor{blue}{\cite{Soh_etal_PRB2021}}. INS measurements were performed at the SEQUOIA spectrometer at the Spallation Neutron Source, Oak Ridge National Laboratory. Three single crystals with a total mass of $\approx 1.8$~g were co-aligned in the $(H, 0, L)$ horizontal scattering plane. The measurements were carried out with incident energies $E_{\textrm{i}}= 50$, $100$, and $150$~meV at $T=5.5(5)$~K by rotating the sample about the vertical axis in $1\deg$ steps over a $270\deg$ range. Throughout the paper, we index the momentum transfer, $\bQ = (H, K, L)$ in reciprocal lattice units (r.l.u) of the $P4/nmm$ lattice, $a=b=4.31(2)$~\AA, $c=10.85(1)$~\AA\ \textcolor{blue}{\cite{Wang_PRM2018,Kealhofer_etal_PRB2018,Qiu2019}}. The data reduction and histogramming to rectangular grid were performed using the MANTID package \textcolor{blue}{\cite{MANTID}} and the MDNorm algorithm \textcolor{blue}{\cite{MDNorm}} (see supplementary information for details \textcolor{blue}{\cite{Supplementary}}).

%***************** Figure 1 *************************************************************************************************
%\begin{figure*}[h]
\begin{figure}
\centering
\includegraphics[width=0.5\textwidth]{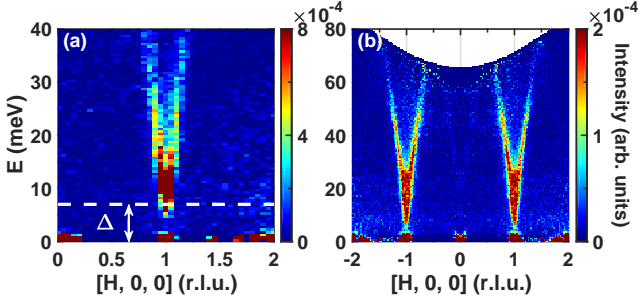}
\caption{{\bf Spin waves in \yms\ in the antiferromagnetic state at $T = 5.5(5)$~K.} Inelastic neutron scattering spectra measured with incident energies $E_\mathrm{i} = 50$ (a) and $100$ (b)~meV showing the dispersion along [$H, 0, 0$] direction. Data bin sizes in $H$ and $K$ are $\pm 0.025$. The data in (a) have bin size $\pm 0.06$ in $L$ and were averaged over $L$ = integers with $L\in[-5,5]$; in (b) were averaged over the continuous interval $L \in [-5, 5]$. The value of $\Delta$ is given in Table~\ref{Tab2ddisp}. For fitting, only the data measured with $E_\mathrm{i} = 100$~meV are used, as shown in Fig~\ref{Fig2:data&fit}. The Gaussian elastic incoherent spectrum obtained by fitting the Q-averaged elastic intensity was subtracted.}
\label{Fig1:gap}
%\end{figure*}
\end{figure}
%**********************************************************************************************************************************************************

\noindent\emph{Results and Analysis.}
Figure~\ref{Fig1:gap} (a),(b) present inelastic neutron scattering spectra for \yms\ in the antiferromagnetic (AFM) phase at $T = 5.5(5)$~K, which reveal the spin wave dispersion along the $\left[H, 0, 0\right]$ symmetry direction. The well-defined spin waves are consistent with the local-moment description and emerge above the AFM wave vector $\bQ_{\rm AFM} = (\pm1, 0, 0)$, as expected for a N\'eel-type magnetic order in \yms\ \textcolor{blue}{\cite{Soh_etal_PRB2021}}. Figure~\ref{Fig1:gap}(a) shows high-resolution data, which clearly demonstrate the presence of a spin-gap, $\Delta \approx 7$~meV, resulting from the uniaxial anisotropy. It also suggests that the spin wave spectrum is slightly blurred along the energy axis, indicating the presence of damping.
The spin-wave dispersion bandwidth along $(H, 0, 0)$, $W = E_\mathrm{\bQ = (1.5,0,0)} \gtrsim 70$~meV, is significantly larger than the values measured in \ymb, \cmb, and \smb\ \textcolor{blue}{\cite{Sapkota2020,Rahn_2017}}, indicating stronger in-plane exchange coupling, $J$. In spin wave theory, $W \sim J$ and $\Delta \sim \sqrt{DJ}$, where $D$ is the uniaxial anisotropy constant. Despite larger $J$, the anisotropy gap in  \yms\ is smaller compared to $\Delta \approx 9$~meV in \ymb\ \textcolor{blue}{\cite{Sapkota2020}}, which is consistent with the weaker SOC of the lighter Sb atoms and hence smaller anisotropy, $D$.
%In AMnBi$_2$ materials, a gap that prevented the formation of the Dirac point and generate massive fermions might be induced by the strong SOC of Bi atoms \textcolor{blue}{\cite{Park_2011,Chaudhuri_2017,Soh2019}}.
%{The bandwidth of the in-plane dispersion along $[H, 0, 0]$ is given by, $W = E_\mathrm{\bQ =(1.5,0,0)} - E_\mathrm{\bQ =(1,0,0)} $ (see below).}

%***********************************   FIGURE 2   ****************************************************************************************
\begin{figure}[b!]
\includegraphics[width=0.5\textwidth]{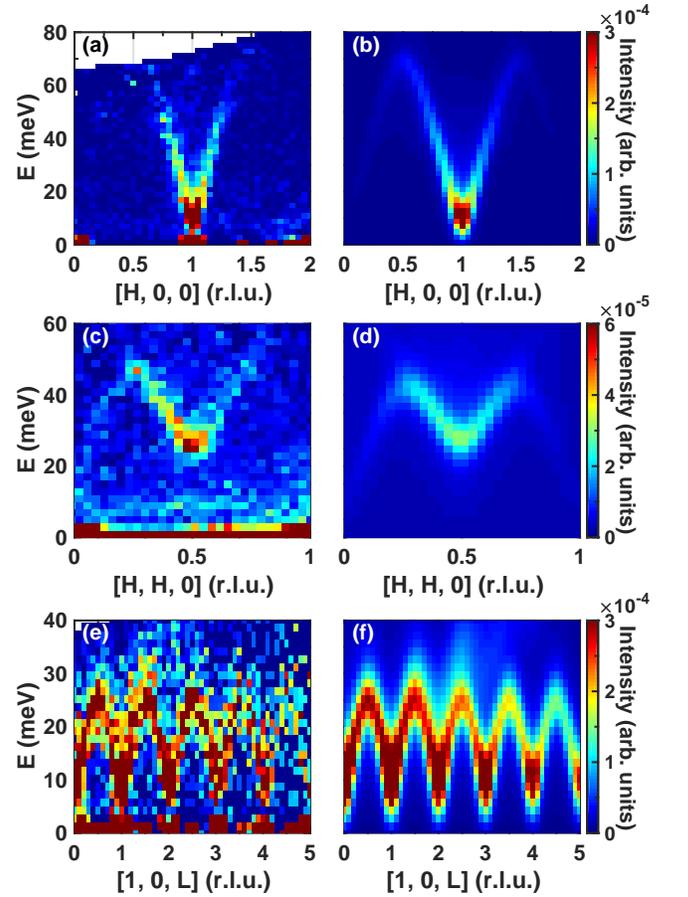}
\caption{{\bf Measured and fitted spin wave spectra of \yms.} The INS spectra measured with $E_\mathrm{i} = 100$~meV at $T = 5.5(5)$ K along three symmetry directions, $[H, 0, 0]$ (a), $[H, H, 0]$ (c), and $[1, 0, L]$ (e). Data bin sizes in (a) are $\pm0.025, \pm0.025, \pm0.06$ in $H$, $K$, $L$, respectively, in (c) are $\pm0.0175, \pm0.035, \pm0.1$ in $(H,H,0)$, $(-H,H,0)$, $L$, respectively, and in (e) are $\pm0.025, \pm0.025, \pm0.05$ in $H$, $K$, $L$, respectively. The spectra in (a) and (c) were averaged for integer $L$ in the range $|L|\leq 5$. (b), (d), and (f) are the INS spectra calculated using Eqs.~\eqref{SDHO} corrected for the instrument resolution and with the fitted parameters listed in Table ~\ref{Tab2ddisp} (see \textcolor{blue}{\cite{Supplementary}} for details).}
\label{Fig2:data&fit}
\end{figure}
%****************************************************************************************************************************************

In order to quantify the interactions and elucidate the presence of damping, we perform quantitative analysis of the measured intensity using an effective spin Hamiltonian, $H = \Sigma_{ij} J_{ij}{\textbf{S}_i}\cdot{\textbf{S}_j} + D\Sigma_{i}(S^z_i)^2$, where $J_{ij}$ includes the interaction between the nearest and next-nearest neighbors in the $ab$ plane ($J_1$ and $J_2$) and nearest neighbors along the $c$ axis ($J_c$). As above, $D$ quantifies the uniaxial anisotropy for the Mn$^{2+}$ spins corresponding to an easy axis along the \textit{c} direction ($D < 0$). In order to account for the spin-wave damping, i.e. the finite spin wave lifetime, we use a damped-harmonic-oscillator (DHO) representation of the dynamical spin correlation function, $S(\bQ, E)$ \cite{Sapkota2020},
\begin{equation}\label{SDHO}
\begin{split}
S(\bq+\bQ_{\rm AFM}, E)\ ={}&
S_{\mathrm{eff}} \dfrac{1}{\pi} \dfrac{2 (A_\mathrm{\bq} - B_\mathrm{\bq})}{1 - e^{- E/k_{\mathrm{B}}T}}\\
%&\times A \dfrac{4 \Gamma E}{\left[E^2\minus E^2(\bq)\right]^2 + 4(\Gamma E)^2} \;.
&\times A \dfrac{\gamma E}{\left[E^2 - E^2_{\bq}\right]^2 + (\gamma E)^2} \;.\end{split}
\end{equation}
Here, $\gamma$ is the damping parameter (Lorentzian FWHM for underdamped DHO), \textit{k}$_{\mathrm{B}}$ is the Boltzmann constant, \textit{S$_{\mathrm{eff}}$} is the effective fluctuating spin, and prefactor $A$ ensures that the DHO spectral function is normalized to 1 (for $(T, \gamma)\rightarrow 0$, $A\rightarrow 1$) \textcolor{blue}{\cite{Supplementary}}. At $T = 5.5(5)$ K $\ll T_N$, spin wave theory gives $A_\mathrm{\bq} = 2S[2J_1 - 2J_2[\sin^2(\pi H) + \sin^2(\pi K)] - 2J_c\sin^2(\pi L) - D]$, $B_\mathrm{\bq} = 4SJ_1 \cos (\pi H) \cos(\pi K)$, and $E^2_{\bq} = A^2_\mathrm{\bq} - B^2_\mathrm{\bq}$.

\begin{figure}[t!]
\includegraphics[width=0.5\textwidth]{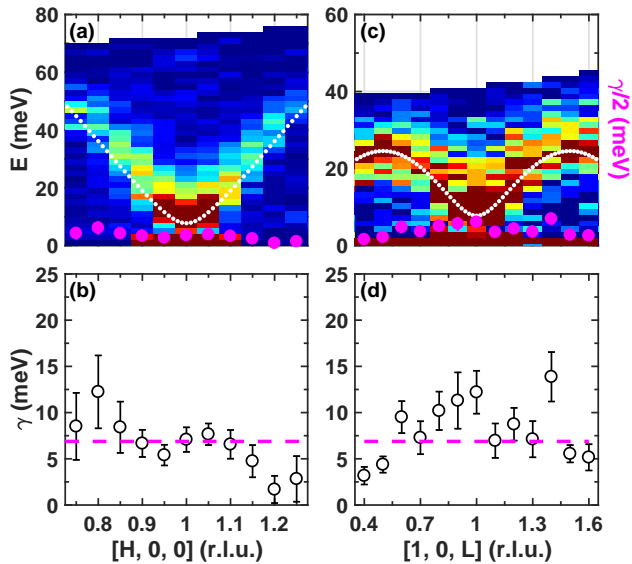}
\caption{{\bf Spin wave dispersion and damping parameter in \yms\ at 5.5(5) K.} The white dotted lines on top of the INS intensity in (a),(c) illustrate the dispersion obtained using the parameters in Table~\ref{Tab2ddisp} without damping. The underdamped spin waves exist where $E_{\bq} > \gamma/2$ (magenta symbols). The symbols show damping obtained by fitting the 1D constant-\bQ\ cuts along $[H, 0, 0]$, (a),(b), and $[1, 0, L]$, (c),(d), directions with the resolution corrected Eq.~\eqref{SDHO}. The magenta dashed lines represent the $\gamma$ value from Table~\ref{Tab2ddisp} obtained from the global 2D fit. Error bars show one standard deviation.}
\label{Fig3:data&multiG}
\end{figure}

We fit the data using Eq.~\eqref{SDHO} convoluted with the instrumental resolution function including the finite $(\bQ, E)$ bin size effects \textcolor{blue}{\cite{Sapkota2020}}. Account for the wave vector resolution is important because the energy line width at each \bQ\ is determined by the convolution, which causes the local averaging over the dispersion \textcolor{blue}{\cite{Supplementary}}. We performed global fits of the 2D energy and wave vector slices shown in Fig.~\ref{Fig2:data&fit}(a),(c),(e) using a single damping parameter, $\gamma$, as well as individual fits of constant-\bQ\ cuts with individual $\gamma(\bQ)$. The INS intensities calculated using the fitted values and the resolution corrected Eq.~\eqref{SDHO} are shown in Fig.~\ref{Fig2:data&fit}(b),(d),(f). The fit results are summarized in Fig.~\ref{Fig3:data&multiG}.

%***********************************   FIGURE 3   ****************************************************************************************
%****************************************************************************************************************************************

The major result of our analysis is the substantial spin wave damping parameter, $\gamma \approx 7.0$~meV, which in \yms\ is nearly twice larger than that in \ymb\ \textcolor{blue}{\cite{Sapkota2020}}. As in \ymb, the damping is roughly \bQ-independent. Figure~\ref{Fig3:data&multiG} shows that the $\gamma$ values obtained by fitting the individual 1D constant-\bQ\ cuts (symbols) fall within about twice the instrumental energy resolution, $E_{res}$, of the 2D global-fitted $\gamma$ value (horizontal dashed line), which closely agrees with the average of $\gamma(\bQ)$. Note, that the absence of $\gamma(\bQ)$ minima near the gap positions, [1, 0, 0] in Fig.~\ref{Fig3:data&multiG}(b) and [1, 0, 1] in Fig.~\ref{Fig3:data&multiG}(d), where the dispersion is flat and \bQ-resolution effects are least important, validates our account for the resolution and corroborates that the observed spin wave broadening is intrinsic. In order to further confirm this, we verified that assuming $\gamma \approx 0$ leads to noticeably inferior quality fits (see Fig.~S3 in \textcolor{blue}{\cite{Supplementary})}.
%We also note that $\gamma$ values extracted from fits along [1, 0, L] direction are overall slightly larger than those along [H, 0, 0] direction, which might indicate some damping anisotropy in this quasi-2Dl material.

%********************************* Table 1 *****************************************************************
\begin{table}[t!]
\caption{\label{Tab2ddisp} Exchange coupling, uniaxial anisotropy, and damping parameters for \yms\ obtained from fitting two-dimensional data
shown in Fig.~\ref{Fig2:data&fit} and those in \ymb\ from Ref. \textcolor{blue}{\onlinecite{Sapkota2020}}.}
\begin{ruledtabular}
\begin{tabular}{ccc}
& \mbox{\ymb}{\textcolor{blue}{\cite{Sapkota2020}}} & \mbox{\yms}\\
\hline
$SJ_1$ (meV) & $25.9\pm0.2$      & $28.1\pm0.1$ \\
$SJ_2$ (meV) & $10.1\pm0.2$      & $10.7\pm0.1$ \\
$SJ_c$ (meV) & $-0.130\pm0.002$ & $-0.597\pm0.023$ \\
$SD$ (meV)   & $-0.20\pm0.01$    & $-0.13\pm0.01$ \\
$\Delta$ (meV) & $9.1\pm0.2$     & $7.7\pm0.4$ \\
$\gamma$ (meV) & $3.6\pm0.2$     & $6.9\pm0.4$ \\

%C\footnote{Some tables require footnotes.}
%  &C\footnote{Some tables need more than one footnote.}
%  & 12537.64 & 37.66345 & 86.37 \\
\end{tabular}
\end{ruledtabular}
\end{table}

%****************************************************************************************************************************************

\noindent\emph{Discussions and Conclusions.}
Understanding the coupling between highly localized magnetic moments of strongly correlated Mn electrons and the Dirac electrons originating in pnictogen (Bi, Sb) layers of $A$/$R$Mn$X_2$ materials presents an important but challenging problem. The layered structure of these systems, where magnetic layers are sandwiched between the layers with the itinerant Dirac electrons, suggests that inter-layer magnetic interactions must involve Dirac fermions. This is further corroborated by observations of a subtle resistivity anomaly at $T_N$ in $A$MnBi$_2$ ($A =$ Ca, Sr) \textcolor{blue}{\cite{Guo_2014}}, indicating a coupling between the Dirac bands and the magnetic ground state. Other studies \textcolor{blue}{\cite{Wang_2012,He2012}}, however, do not report the anomaly. Similarly, no evidence that the magnetic dynamics are influenced by the Dirac/Weyl fermions was obtained from the spin wave analyses of the INS measurement of magnetic excitations which did not consider spin wave damping \textcolor{blue}{\cite{Rahn_2017,Soh2019,Cai}}.

The reason for the difficulty of experimentally observing the manifestations of spin-fermion coupling with Dirac electrons is that the linear Dirac dispersion has a low density of states and therefore their effect on spin wave excitations is weak. Nevertheless, a thorough analysis of spin wave spectra measured by INS in \ymb, similar to the one presented here, did find a non-negligible spin wave damping, $\gamma \approx 3.6$~meV (Table~\ref{Tab2ddisp}) \textcolor{blue}{\cite{Sapkota2020}}. A comparison with the theoretical model showed that albeit small, this damping is a signature of a very substantial spin-fermion coupling.

The results of our analysis presented in Table~\ref{Tab2ddisp} show that the damping parameter in \yms\ is about twice larger than that in \ymb, while the inter-layer interaction is roughly four times larger in magnitude and the intra-layer interaction $J_1$ is $\sim10 \%$ larger. In addition to establishing experimental systematics, these quantitative relationships suggest that Dirac charge carriers may in fact participate in mediating all magnetic interactions between Mn moments, both intra- and inter-plane. In this scenario, it might be instructive to infer functional relationships between $J_{1,2}, J_c$, and $\gamma$, such as $J_c \propto \gamma^2$, which comply with the experimental observations and provide experimental guidance for future theories.

In summary, our INS measurements of magnetic excitations in single crystals of Dirac semimetal \yms\ reveal considerable broadening of the antiferromagnetic spin waves at low temperature, $T \approx 5.5$~K$ \ll T_N$, which is consistent with substantial spin-fermion coupling in this material. By fitting the measured spin wave spectra to Heisenberg model with easy-axis anisotropy and with finite spin wave lifetime (damping), we extracted the damping parameter, $\gamma = 6.9(4)$~meV, and inter- and intra-layer exchange interactions. Comparison of the obtained model parameters with those in \ymb\ and other 112 Dirac materials allows establishing systematic phenomenology of spin-fermion coupling in these systems and suggests that Dirac electrons are involved in the inter-layer spin coupling and might also participate in all magnetic interactions between Mn$^{2+}$ ions. While developing theoretical description of such an RKKY-type coupling via Dirac electrons presents a challenge for the future, our results provide experimental guidance for such theories and an input for predictive theory of the magnetotransport phenomena in this regime.

%\end{Discussions and Conclusions}

\noindent\emph{Note added.}
While this work was being finalized for publication, a related INS study \textcolor{blue}{\cite{Tobin_arXiv2023}} of \yms\ using triple axis spectroscopy (TAS) appeared. While constraints of instrumental resolution ($\Delta E_{FWHM} \approx 8$~meV) inherent to TAS in the energy range relevant for this study did not allow those authors to explore spin wave damping and resulted in moderately different Hamiltonian parameters (refined by fitting triple axis measurements to the same model as we use here but without damping), the general trends and conclusions reported in \textcolor{blue}{\cite{Tobin_arXiv2023}} support our results. In particular, they support the conclusion that spin-fermion coupling in \yms\ is stronger and more important compared to other 112 systems. It is also noteworthy that half-polarized neutron diffraction reported in \textcolor{blue}{\cite{Tobin_arXiv2023}} confirms the localized, ionic nature of Mn magnetic moments, giving direct experimental support to models of spin-fermion coupling such as proposed in our earlier work \textcolor{blue}{\cite{Sapkota2020}}.

{\bf{Acknowledgements}} We gratefully acknowledge discussions with A. Tsvelik and technical assistance from V. Fanelli. This work at Brookhaven National Laboratory was supported by Office of Basic Energy Sciences (BES), Division of Materials Sciences and Engineering,  U.S. Department of Energy (DOE), under contract DE-SC0012704. This research used resources at the Spallation Neutron Source, a DOE Office of Science User Facility operated by the Oak Ridge National Laboratory.

%
%{\bf{Author contributions: }}%\\
%I.A.Z. conceived and directed the study. I.A.Z., A.S. and J.M.T. designed the study; Z.X. and C.P. provided the samples for the study; I.A.Z. and A.I.K. carried out neutron experiments and obtained the neutron data; I.A.Z., A.T.S. and X.H. reduced the data for analysis; X.H. and I.A.Z. analyzed the data, prepared the figures and wrote the paper, with input from all authors.
%%
%{\bf{Competing Interests: }}%\\
%The authors declare that they have no competing interests.
%%
%{\bf{Data availability: }}%\\
%All data needed to evaluate the conclusions in the paper are present in the paper and/or the Supplementary Materials. Additional data available from authors upon reasonable request.

%\bibliography{BibTeX_YbMnSb2}
%\end{document}

%\end{document}

\begin{widetext}
\pagebreak
\hypersetup{pageanchor=false}
\renewcommand{\thepage}{S\arabic{page}}
\setcounter{page}{1}
\renewcommand{\theequation}{S\arabic{equation}}
\setcounter{equation}{0}
\renewcommand{\thefigure}{S\arabic{figure}}
\setcounter{figure}{0}

\section*{Supplementary Information}

\begin{center}
{\bf Coupling of magnetism and Dirac fermions in \yms\ } \\

\bigskip
Xiao Hu, Aashish Sapkota, Zhixiang Hu, Andrei T. Savici, Alexander I. Kolesnikov, John M. Tranquada, Cedomir Petrovic, and Igor A. Zaliznyak \\
\bigskip
correspondence to: zaliznyak@bnl.gov
\end{center}

\bigskip
\noindent{\bf This PDF file includes:}\\
Supplementary Text\\
Supplementary Figures S1-S7\\
%Supplementary Theory\\
%Supplementary References\\
%\newpage

\subsection{Account for the resolution}
\label{Resolution}
In order to compare the measured energy- and wave-vector-dependent inelastic neutron scattering (INS) intensity with the model dynamical spin correlation function describing damped spin wave response, $S(\bQ, E)$ given by Eq.~(1) of the main text, we need to, (i) convert $S(\bQ, E)$ to magnetic neutron scattering cross-section, $\Sigma(\bQ, E)$ \cite{ZaliznyakLee_MNSChapter}, and (ii) account for the instrumental resolution effects. Up to an arbitrary normalization multiplier, the measured neutron intensity shown in the main text has been reduced \cite{MANTID,MDNorm} so that it is given by the part of magnetic INS cross-section consisting of the product of $S(\bQ, E)$ and the square of magnetic form factor of the unpaired electrons. We find that magnetic form factor of Mn$^{2+}$ ion adequately describes our data, so we use it in our analysis and fitting.

The instrumental resolution effects are described by the convolution,
\begin{equation}
\label{resolution}
I(\bQ, E) = \int_{-\infty}^{\infty}{R(\bQ', E')\Sigma(\bQ - \bQ', E - E')}d\bQ' dE' ,
\end{equation}
accounting for the probability that an instrument measures cross-section offset from the nominal $\bQ$ and $E$ by amounts $\bQ'$ and $E'$, respectively, with the probability distribution described by the resolution function, $R(\bQ', E')$. For the time-of-flight (TOF) spectrometers such as SEQUOIA used in our work, the resolution in $\bQ$ and energy are, to a good approximation, uncoupled, resulting in a factorized resolution function, $R(\bQ, E) = R_{\bQ}(\bQ) R_{E}(E)$ \cite{Abernathy_2012}. Binning to hyper-rectangular bins (voxels) whose boundaries along the energy direction are independent of $\bQ$ preserves such decoupling. Hence, we consider the effects of energy and wave vector resolution separately.

\subsubsection{Account for the energy resolution}
\label{E-resolution}
In order to fully account for the resolution effects, one needs to consider both linewidth broadening by the instrument energy resolution and the finite energy bin size effect at each \textbf{Q}. The TOF instrument energy resolution is approximated by a Gaussian function, which can be calculated knowing the instrument parameters \cite{Abernathy_2012}, and for zero energy transfer can also be measured through incoherent elastic scattering. Comparing the result of the calculation and the measurement provides a consistency and accuracy check for the calculation. In our measurements, the full width at half maximum (FWHM) of the fitted Gaussian incoherent elastic peak, which was subtracted from the data, was $\Delta E = 1.41$~meV for $E_i = 50$~meV (calculated $\Delta E = 1.37$~meV), $\Delta E = 2.51$~meV for $E_i = 100$~meV (calculated $\Delta E = 2.60$~meV), and  $\Delta E = 3.68$~meV for $E_i = 150$~meV (calculated $\Delta E = 3.97$~meV). The measured values are in good agreement with the calculated DGS energy resolution for our setup (\href{https://rez.mcvine.ornl.gov}{https://rez.mcvine.ornl.gov}, \cite{Abernathy_2012}). %which are $\Delta E = 1.37$~meV for $E_i = 50$~meV, $\Delta E = 2.60$~meV for $E_i = 100$~meV, and  $\Delta E = 3.97$~meV for $E_i = 150$~meV,

%***************** Figure S1 ********************************************************************************
%\begin{figure*}[h]
\renewcommand{\thefigure}{S\arabic{figure}}
\begin{figure}
%\centering
%\includegraphics[width=1.\textwidth]{Figures/2ddisp_3Ei_data.pdf}
%\includegraphics[width=0.5\textwidth,left]{Fig1_gap&data_vector.png}
\includegraphics[width=0.5\textwidth]{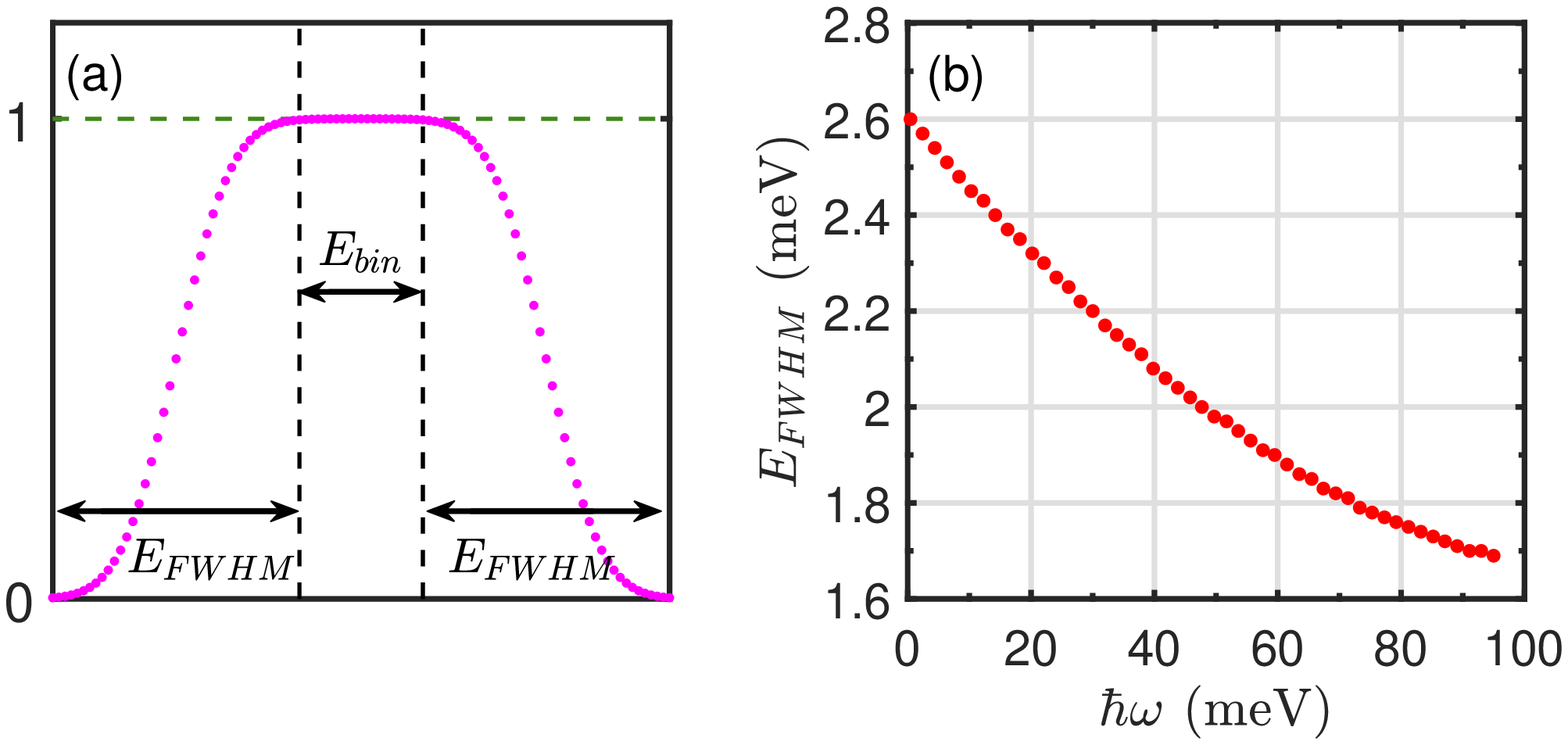}
\caption{(a) Illustration of the energy resolution weight function used in our fits. $E_{bin}$ is the actual data bin size used to create plots and $E_{FWHM}$ is the energy resolution of SEQUOIA with our experimental settings. The spin-wave intensity at each $E$ was calculated as a weighted average of intensities within the interval of size $E_{bin}+2E_{FWHM}$ weighted by the normalized resolution weight function centered at this $E$. (b) SEQUOIA energy resolution, $E_{FWHM}$, as a function of energy transfer, $\hbar\omega$, for $E_{i}$ = 100 meV and high-resolution Fermi chopper frequency of 600 Hz used in our measurements. The energy resolution is obtained from \href{https://rez.mcvine.ornl.gov}{https://rez.mcvine.ornl.gov}.}
\label{supp1}
%\end{figure*}
\end{figure}
%**********************************************************************************************************************************************************

The effect of binning amounts to a convolution of the instrument Gaussian energy resolution, $R^{(0)}_E(E)$, with the window function corresponding to the energy interval used for binning,
\begin{equation}
\label{E-resolution}
R_{E}(E) = \int_{-E_{bin}/2}^{E_{bin}/2}{R^{(0)}_E(E-E')} dE' =  \int_{0}^{E+E_{bin}/2}{R^{(0)}_E(E')} dE' - \int_{0}^{E-E_{bin}/2}{R^{(0)}_E(E')} dE'.
\end{equation}
The resulting energy resolution function is simply a difference of the two complementary error functions parameterized by the FWHM of the instrument energy resolution, $E_{FWHM}$, and the energy binning size, $E_{bin}$, as shown in Fig.~\ref{supp1}. The energy bin sizes we used for histogramming the data are $E_{bin}$ = 1.5 meV for dispersion along $[1, 0, L]$ direction and $E_{bin}$ = 2.0 meV for $[H, 0, 0]$ and $[H, H, 0]$ directions and for constant-$E$, $(H,K)$ data slices.

\subsubsection{Account for the wave vector resolution}
\label{Q-resolution}

%***************** Figure S0 ********************************************************************************
%\begin{figure*}[h]
\renewcommand{\thefigure}{S\arabic{figure}}
\begin{figure}
%\centering
%\includegraphics[width=1.\textwidth]{Figures/2ddisp_3Ei_data.pdf}
%\includegraphics[width=0.5\textwidth,left]{Fig1_gap&data_vector.png}
\includegraphics[width=0.5\textwidth]{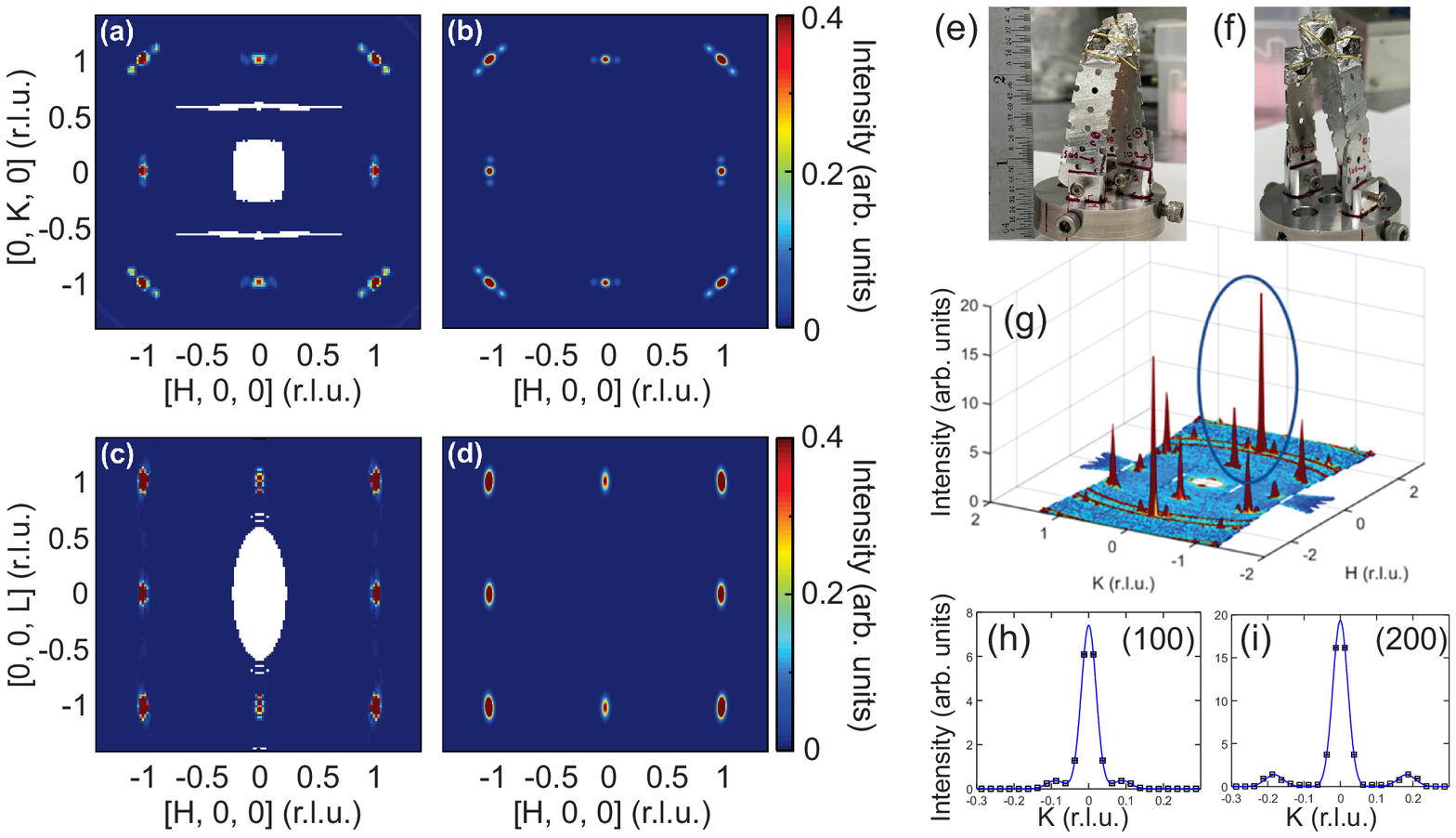}
\caption{(a),(c) Color contour maps of elastic scattering from our YbMnSb$_2$ sample measured with $E_i = 100$~meV in $[H, K, 0]$ ($|L| \leq 0.1$~r.l.u.) and $[H, 0, L]$ ($|K| \leq 0.025$) reciprocal planes, respectively; the binning along the X and Y axes of the plot is $\pm 0.01$~r.l.u. The data was averaged within energy bin $|E|<1$~meV and symmetrized with respect to $H \leftrightarrow -H$, $K \leftrightarrow -K$, $L \leftrightarrow -L$ (we checked that symmetrization introduces no spurious features by comparatively inspecting the non-symmetrized data). (b), (d) 2D Gaussian fits to the data in (a), (c), respectively, quantifying the instrument wave vector resolution. The peak is nearly isotropic in the $HK$ plane, with $Q_{FWHM}^{\perp} = 0.049$~r.l.u. and $Q_{FWHM}^{\|} = 0.041$~r.l.u., indicating small crystal mosaic which contributes very little to the $\bQ$ resolution [see also panels (g)-(i)]. In our analysis, we use the average values, $Q^{H,K}_{FWHM} = 0.045$~r.l.u. and $Q^{L}_{FWHM} = 0.1$~r.l.u. (e),(f) Photographs of the 3-crystal YbMnSb$_2$ sample used in our measurements. (g) An extended coverage 3D surface plot of elastic intensity measured with $E_i = 50$~meV for $|E| < 0.5$~meV and $|L| < 0.1$~r.l.u., binned with $\pm 0.125$~r.l.u. along $H$ and $K$; the oval highlights $(100)$ and $(200)$ Bragg peaks for which line scans with Gaussian fits are shown in panels (h) and (i), respectively. The obtained peak width, $Q^{\perp,(100)}_{FWHM} = 0.047$~r.l.u. and $Q^{\perp,(200)}_{FWHM} = 0.049$~r.l.u. does not scale with the wave vector length, $Q$, indicating that peak width is not mosaic-limited (the distance to small satellite peaks coming from minor crystallites misaligned by $\approx 6^{\circ}$ and amounting to $\sim 7\%$ of the main peak does scale with $Q$). From these fits, an upper limit on the crystal mosaic, $\eta < 1^{\circ}$, can be inferred. The follow-up scans with a triple axis spectrometer indicate $\eta \lesssim 0.5^{\circ}$. }
\label{supp0}
%\end{figure*}
\end{figure}
%**********************************************************************************************************************************************************

Similar to the energy resolution, the wave vector resolution function is a convolution of the Gaussian instrumental resolution, accounting also for the sample mosaic, with the three dimensional (3D) window function corresponding to our $\bQ$-binning. In order to minimize the effect of wave vector resolution and still have sufficient intensity for reliable fitting, we used the binning size of $(\pm0.025$, $\pm0.025$, $\pm0.06)$ in $(H, K, L)$, respectively, for $[H, 0, 0]$ data, $(\pm0.025$, $\pm0.025$, $\pm0.05)$ in $(H, K, L)$, respectively, for $[1, 0, L]$ data, $(\pm0.0175, \pm0.035, \pm0.1)$ in $([H,H,0], [-H,H,0], L)$, respectively, for $[H, H, L]$ data, and $(\pm0.025, \pm0.025, \pm0.1)$ in $(H, K ,L)$ for constant-($L, E$) data. The instrumental $\bQ$-resolution FWHM, $Q_{FWHM}$, was obtained by fitting nuclear elastic Bragg peaks binned on a much finer grid, $\pm 0.01$ in $[H, K, 0]$ and $[H, 0, L]$ reciprocal lattice planes, to Gaussian functions (Fig.~\ref{supp0}).

The resolution function ranges used for numerical convolution were $\Delta_{\alpha}=(Q_{bin})_{\alpha}+2(Q_{FWHM})_{\alpha}$, where $\alpha$ indexes the $H, K, L$ directions. For each $\bQ$, a 3D reciprocal mesh block with the size of $\Delta_{H} \times \Delta_{K} \times \Delta_{L}$ was generated, centered at this $\bQ$. The spin wave intensity at each $\bQ$ point was calculated as a weighted average of intensities within the mesh block, weighted by the 3D resolution function similar to that in Fig.~\ref{supp1}. To ensure the accuracy of the calculation, the mesh grid ($n_{H} \times n_{K} \times n_{L}$, where $n_{i}$ is the number of points used along each direction within the 3D mesh block in the numerical calculation) was chosen to be $11\times 11\times 9$. Fig.\ref{supp2} shows that in the absence of any intrinsic line width, the binning size $\delta H = \pm0.02$ and $\delta K = \pm0.025$ would introduce a peak width of $\sim 7 - 8$ meV along $[H, 0, 0]$. This is comparable to the values of $\gamma$ obtained in our fits and therefore it is important to account for this artifact, which exists in the data due to non-negligible binning-size effects, when fitting data to theoretical model.

%***************** Figure S2 ********************************************************************************
%\begin{figure*}[h]
\renewcommand{\thefigure}{S\arabic{figure}}
\begin{figure}[h]
%\centering
%\includegraphics[width=1.\textwidth]{Figures/2ddisp_3Ei_data.pdf}
%\includegraphics[width=0.5\textwidth,left]{Fig1_gap&data_vector.png}
\includegraphics[width=0.5\textwidth]{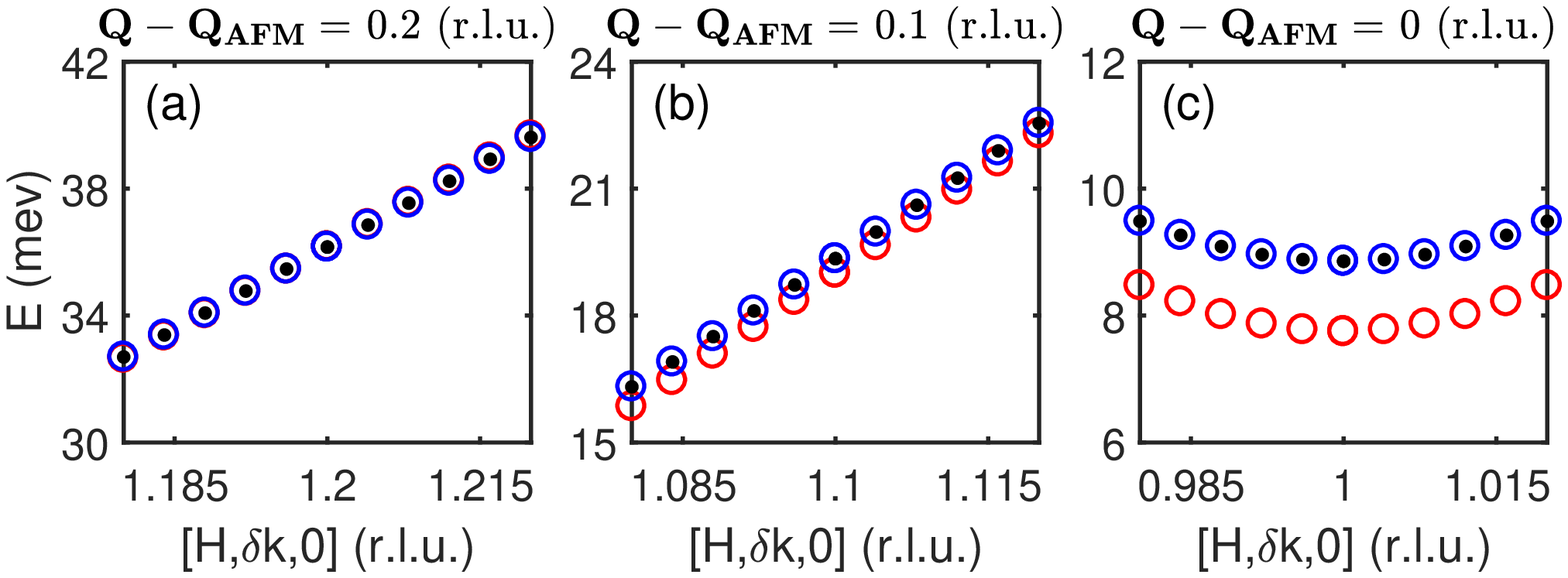}
\caption{Illustration of the energy range spanned by the spin wave dispersion in YbMnSb$_2$ within finite bin size at different $\bQ$ along $[H, 0, 0]$. Markers are the $E$ values obtained using the Hamiltonian parameters from our fits listed in Table~1 of the main text for $\bQ = (H, \pm \delta k 0, 0)$ around $H = 1.2$ (a), $H = 1.1$ (b), $H = 1.0$ (c). Red open circles are for $\delta k = 0$, blue open circles and black closed circles are for $\delta k = \pm 0.025$~r.l.u., respectively.}
\label{supp2}
%\end{figure*}
\end{figure}
%**********************************************************************************************************************************************************

\subsection{Details of the fitting procedure}
\label{Fitting}

\subsubsection{Account for multi-crystal sample and ``shadow dispersion''}

%***************** Figure S4 ********************************************************************************
%\begin{figure*}[h]
%\renewcommand{\thefigure}{S\arabic{figure}}
\begin{figure}[h]
%\centering
\includegraphics[width=0.5\textwidth]{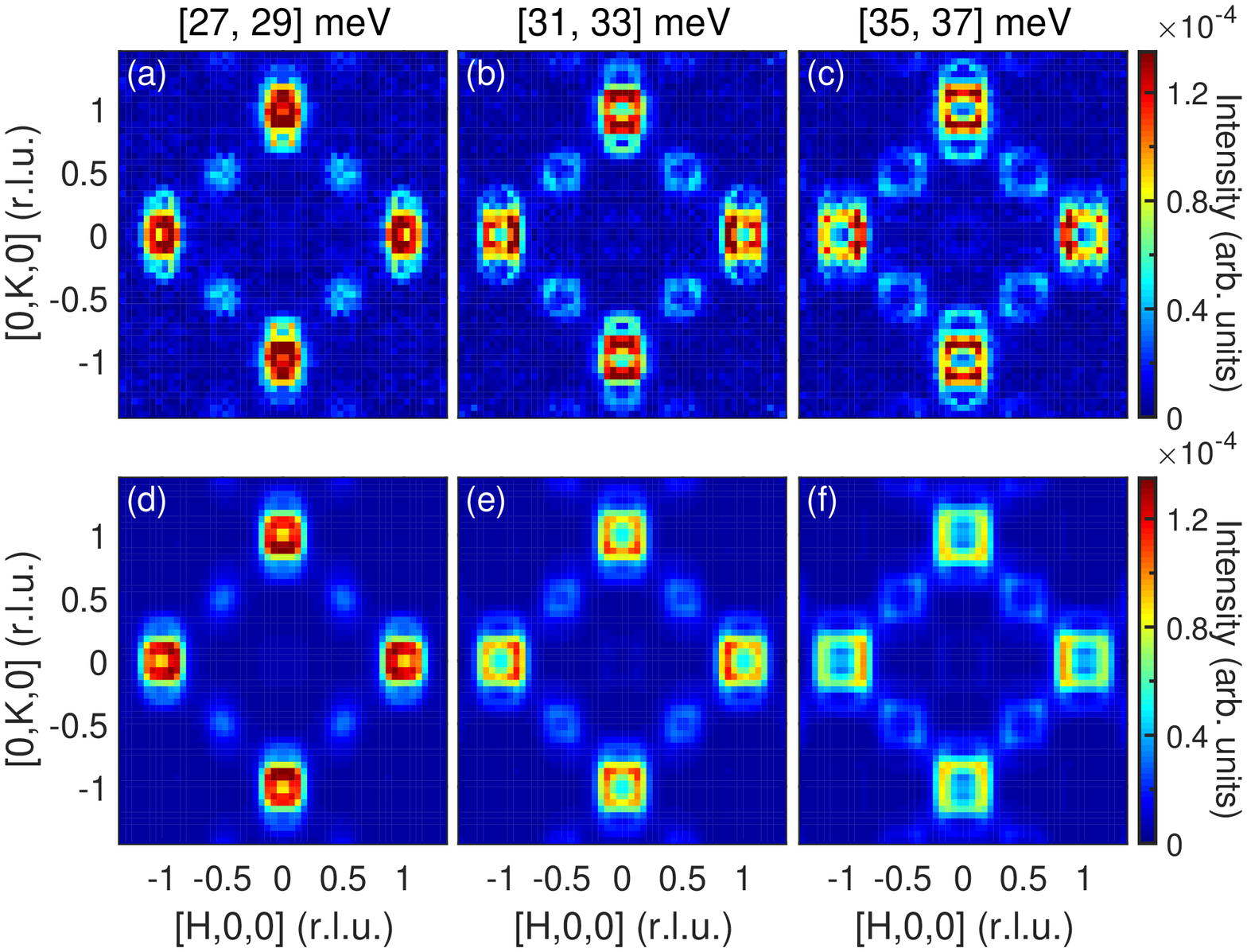}
\caption{(a) - (c) Selected constant - $(L, E)$ data slices with $|L|\in[2.5, 3.5]$ revealing the shadow dispersion from misaligned crystals, (d) - (f) is the corresponding calculated intensity. The shadow dispersion is adequately reproduced by the optimized model accounting for the angular offsets of the two crystals.}
\label{supp4}
%\end{figure*}
\end{figure}

Our sample consists of three co-mounted single crystals (Fig.~\ref{supp0}), each with mosaic $\eta \lesssim 0.5^\circ$ (as measured in a follow-up alignment experiment on a triple axis spectrometer). The INS measurements presented in the manuscript were performed in remote mode during the pandemic lockdown and, as a result, crystals alignment was not adjusted upon shipping. Consequently, samples were slightly misaligned, which is seen in an appearance of the ``shadow dispersion'' in constant-energy slices (Fig.~\ref{supp4}). Figure~\ref{supp4} (a)–(c) present selected constant-$(L, E)$ slices with $|L| \in [2.5, 3.5]$. Along with the main spin-wave dispersion, an additional intensity due to sample misalignment is also observed on a ``shadow dispersion'' displaced along $K$ direction.

In order to account for the misalignment, we carried out careful analysis of the measured spin wave intensity accounting for the three contributions coming from the three crystals in our sample. We parameterized the misaligned crystals by their offset angles, ($\theta_H, \theta_K, \theta_L$), and the relative masses, $m_i$ ($i=1,2$ indexes the misaligned crystals), with respect to the main, aligned crystal. Fitting the observed ``shadow dispersion'' allowed us to refine the parameters of the misaligned crystals. We found that the misalignment is mainly via rotation around the $a$-axis ($H$ direction) of the main crystal, while misalignment with respect to $b$- and $c$-axis is negligible. Using the misalignment angles and mass ratios of misaligned sample pieces relative to the main piece as fitting parameters, we obtain  $\theta_{H}= 9.4(2)^{\circ}, \theta_K = 0.1(1)^{\circ}, \theta_L= 0.4(4)^{\circ}$, and mass ratios $m_1 = m_2 = 0.3(1)$. Figure~\ref{supp4} (d)-(f) show the calculated constant-$(L, E)$ slices with the optimized parameters. Figure~\ref{supp5} presents selected constant-$H$ cuts from slices in Figure~\ref{supp4}, with $H\in[-0.3,0.3], [0.3, 0.75], [0.75,1.3]$. We observe that the model with the optimized parameters cited above adequately reproduces both the main spin-wave intensity and the shadow dispersion, confirming the refined relative crystal alignment in our sample.

%sample pieces relative to the major piece, because of which we use $mr_1$ and $mr_2$ to represent the mis-aligned sample mass ratios and ($\theta_H, \theta_K, \theta_L$) for misalignment of sample $m_1$, ($-\theta_H, -\theta_K, -\theta_L$) for sample $m_2$, respectively. Here, $\theta_H, \theta_K, \theta_L$ represent the mis-aligned angle about \textit{a}, \textit{b}, \textit{c}-axis, respectively.

%***************** Figure S5 ********************************************************************************
%\begin{figure*}[h]
%\renewcommand{\thefigure}{S\arabic{figure}}
\begin{figure}[h]
%\centering
\includegraphics[width=0.5\textwidth]{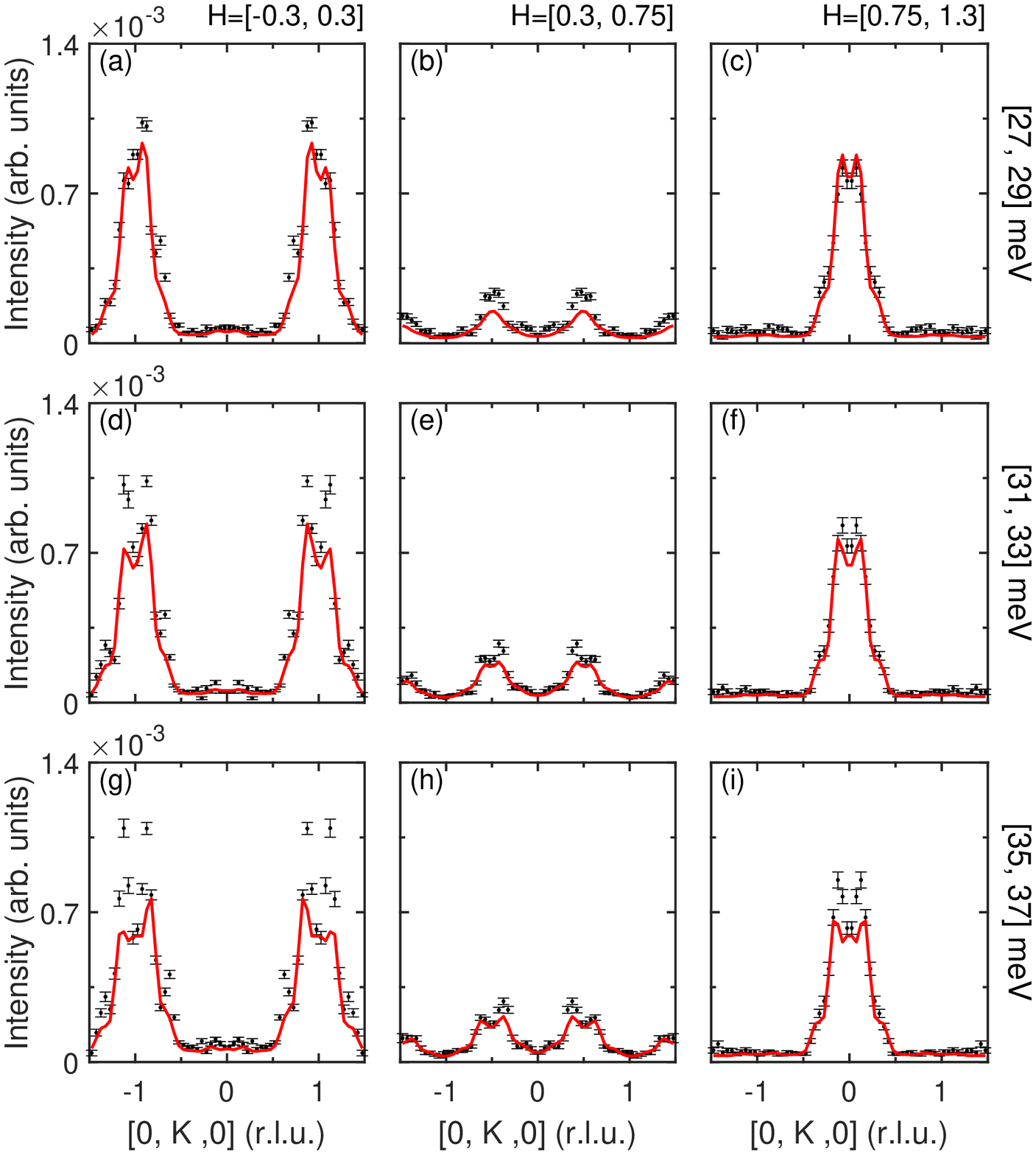}
\caption{Selected constant-$H$ cuts of the data (symbols) and fitting result (solid lines) for $|L|\in[2.5, 3.5]$ and $E\in[27, 29]$~meV, (a)-(c), $[31, 33]$~meV, (d)-(f), and $[35, 37]$~meV, (g)-(i). }
\label{supp5}
%\end{figure*}
\end{figure}

\subsubsection{Data sets used for fitting}
We used two sets of data slices in our model fitting, $Q-E$ slices corresponding to the dispersion spectra along $[H, 0, 0]$, $[1, 0, L]$, and $[H, H, 0]$ directions (Fig.~2 of the main text) and $Q-Q$ slices corresponding to constant-($L, E$) slices with $E = [27, 29]$~meV, $[31, 33]$~meV, and $[35, 37]$~meV and $|L| \in [2.5, 3.5]$ (Figs~\ref{supp4},\ref{supp5}). The $E$ and $L$ intervals used for $Q-Q$ slices were chosen to optimize statistics of the ``shadow dispersion'' pattern. In fitting, the statistical weight of the three $Q-E$ spectra has been adjusted so that they contribute to the reduced chi-squared on par with $Q-Q$ data. For $Q-E$ data, only the data at $E >$ 4 meV were used in fitting to avoid data contamination from the incoherent elastic background and magnetic Bragg peaks. The fits were carried out using the magnetic form factor of $\text{Mn}^{2+}$. The $[H, 0, 0]$ and $[H, H, 0]$ spectra were obtained by averaging slices with \textit{L} = integers in the range of $L \in [-5, 5]$ and the fitted intensity was calculated by averaging the corresponding calculated contributions.

%***************** Figure S6 ********************************************************************************
\renewcommand{\thefigure}{S\arabic{figure}}
\begin{figure}[h!]
%\centering
\includegraphics[width=0.5\textwidth]{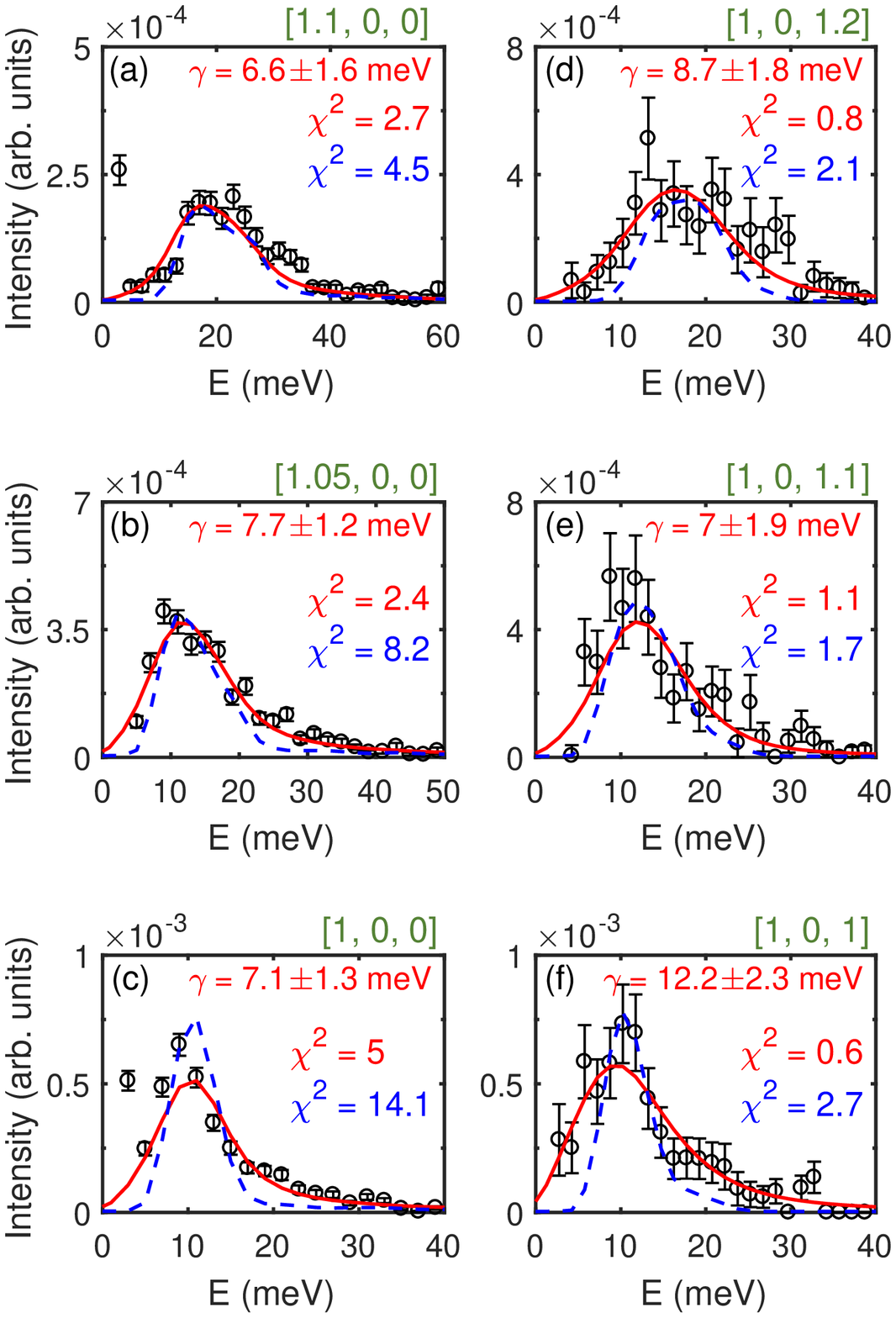}
\caption{Selected resolution-corrected fits of the one-dimensional constant-$\bQ$ cuts of $E_i = 100$~meV, 5.5 K data. (a) – (c) cuts along the $[H, 0, 0]$ direction and (d) – (f) along $[1, 0, L]$ direction. Data are averaged over the range of $\pm 0.025$ in $H$ and $K$ and $\pm 0.06$ in $L$. Red solid lines are the DHO fits where $\gamma(\bQ)$ was fitted (Fig.~3 of the main text) and blue dashed lines are fits with fixed $\gamma$ = 0.11 meV; in both cases peak position was adjusted by varying $SJ_1$ and $SJ_2$ as described in section \ref{Fitting}.}
\label{supp6}
%\end{figure*}
\end{figure}
%************************************************************************************************************

%***************** Figure S3 ********************************************************************************
\renewcommand{\thefigure}{S\arabic{figure}}
\begin{figure}[h!]
%\centering
\includegraphics[width=0.5\textwidth]{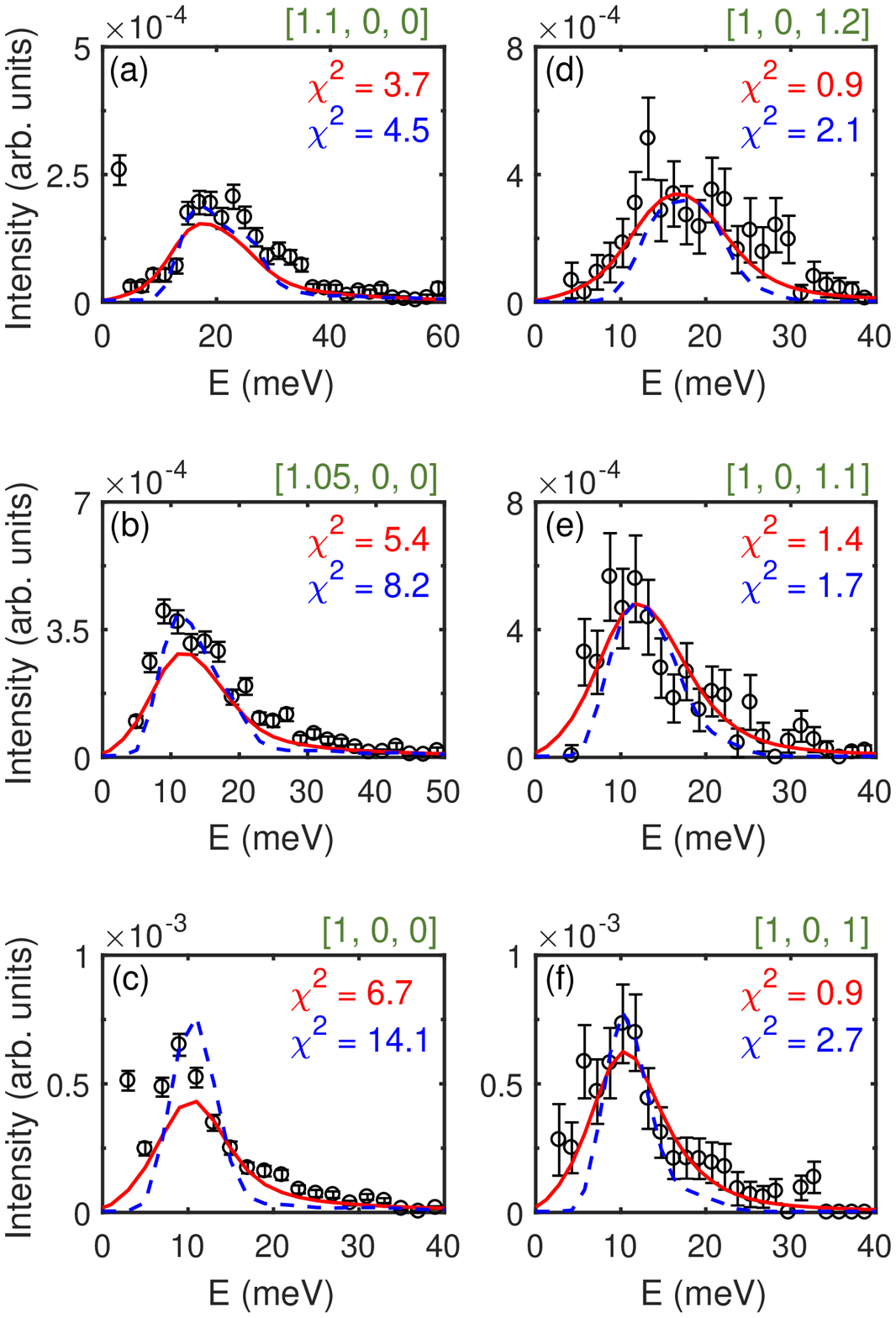}
\caption{Selected one-dimensional constant-$\bQ$ cuts of $E_i = 100$~meV, 5.5 K data and of the corresponding resolution corrected 2D global fits. (a) – (c) are cuts along the $[H, 0, 0]$ direction and (d) – (f) are cuts along $[1, 0, L]$ direction. Data are averaged over the range of $\pm 0.025$ in $H$ and $K$ and $\pm 0.06$ in $L$. Red solid lines are the DHO fits where $\gamma = 6.9(4)$ was fitted and blue dashed lines are fits with fixed $\gamma$ = 0.11 meV.}
\label{supp3}
%\end{figure*}
\end{figure}
%**********************************************************************************************************************************************************

\subsubsection{Recursive fitting}
$Q-E$ dispersion data are sensitive to interaction parameters, uniaxial anisotropy, and spin-wave damping (in energy) $(SJ, SD, \gamma)$ and are weakly sensitive, if at all, to misaligned crystals. On the other hand, $Q-Q$ data are also sensitive to sample misalignment angles and misaligned sample mass $(\theta, m)$. Therefore, we fitted the two data sets in a recursive manner, where we first fitted the $Q-E$ data ($[H, 0, 0]$ and $[H, H, 0]$ dispersion spectra) to obtain an estimate of the interaction parameters. In fact, the in-plane nearest-neighbor interaction, $SJ_1$, can already be estimated from the spin-wave dispersion, $4(SJ_1)^2  = E^2_{(1.5, 0, 0)} - E^2_{(0.25, 0.25, 0)}$, using values of $E_{(1.5, 0, 0)}$ and $E_{(0.25, 0.25, 0)}$ gauged from the raw data. Inspection of spin-wave spectra along $[H, 0, 0]$ and $[H, H, 0]$ directions gives $E_{(1.5, 0, 0)}\approx 73 $~meV and $E_{(0.25, 0.25, 0)}\approx 47$~meV, which yields $SJ_1 \approx 28$~meV.

$Q-Q$ data around $(H, K)=(0,\pm1)$, $|L|\in[2.5, 3.5]$, (Fig.~\ref{supp4}) reveal no shadow dispersion offset in $H$ direction, suggesting negligible misalignment ($\theta_K, \theta_L$) about $b$- and $c$-axis. On the other hand, clear ``shadow dispersion'' intensity is observed offset along $K$ around $(H, K)=(\pm1, 0), (0, \pm1)$, suggesting a contribution form the crystal misaligned about $a$-axis, ($\theta_H$). From the ratio of intensities of the ``shadow dispersion'' offset along $K$ and the principal dispersion at $(H, K)=(1, 0)$ is straightforward to estimate the mass of the misaligned sample pieces, $m_1\approx m_2 \approx 0.3$ and the misalignment angle, $\theta_H \approx 9^{\circ}$, while within the fitting error, $\theta_K \approx \theta_L \approx 0$.

In our recursive fitting procedure, we first fix the $SJ_1, m_1, m_2, \theta_i$ parameters to their previously refined values and fit the $Q-E$ data varying the remaining parameters, $(SJ_2, SJ_c, SD, \gamma)$ (sample misalignment parameters, $m_1, m_2, \theta_i$, have minor effect on the $Q-E$ data fitting because the constant-$E$ slices in $[H, 0, 0], [H, H, 0], [1, 0, L]$ dispersion directions do not contain significant contribution of the ``shadow dispersion''). We then fix $(SJ_2, SJ_3, D, \gamma)$ to the fitted values and fit $SJ_1$. We continue iteration until the convergence is reached and parameter changes are smaller than the error bars. Then, the optimized interaction parameters, uniaxial anisotropy, and damping obtained from this recursive fitting are fixed to their  fitted values and the sample misalignment parameters, $m_1, m_2, \theta_i$, are refined by fitting the $Q-Q$ data (constant-($L, E$) slices). Then, we fix the misalignment angles and misaligned sample mass to their fitted values and again fit the $Q-E$ data in the recursive way described above. We repeat this iteration until convergence is reached and parameter changes are smaller than the error bars. Given the large amount of data points we fit, the overall quality of the fit, $\chi^2 \approx 3$, indicates that our fitting results are reliable.
%The criterion of convergence mentioned above is that changes of the fitted parameter values are smaller than their error bars.

\subsubsection{\bQ-dependence of the damping parameter}
To investigate the $\bQ$-dependence of the damping parameter, $\gamma(\bQ)$, we fitted one-dimensional constant-$\bQ$ cuts extracted from $[H, 0, 0]$ and $[1, 0, L]$ spectra using Eq.~(1) of the main text (Fig.~\ref{supp6}). The fits were done accounting for the instrumental resolution and bin sizes, as discussed in section \ref{Resolution} above. In order to avoid biased fitting of $\gamma(\bQ)$ at $\bQ$ points around $H = 0.75$ or $1.25$ in $[H, 0, 0]$ spectra where the optimized global 2D fit has slight discrepancies in dispersion energy ($\lesssim 2E_{bin}$, cf Fig.~2 of the main text), we allow $SJ_1$ and $SJ_2$ to vary within a small interval, $[SJ_{1,2} - E_{bin}/2, SJ_{1,2} + E_{bin}/2]$, determined by the energy bin size, $E_{bin}=2$~meV, around their values obtained from the 2D global fit when fitting these cuts. This allows to adjust the peak position and improve the fit quality of these $\bQ$ cuts, so the corresponding $\gamma(\bQ)$ is more reliably evaluated (Fig.~\ref{supp6}). The reason for varying only $SJ_1$ and $SJ_2$ is that $[H, 0, 0]$ spectra are insensitive to $SJ_c$, and the small $SD$ influences the spin gap at $H=1$, but not $\bQ$ points away from the gap. The obtained $\gamma(\bQ)$ values are shown in Fig.~3 of the main text.

\subsubsection{Manifestation of the intrinsic physical width}
Figures \ref{supp6} (a) – (f) present selected constant-$\bQ$ cuts of 5.5 K data with 1D fits done with and without the intrinsic damping line width, $\gamma$, and accounting for the instrumental resolution and bin sizes as described above. The results demonstrate the non-negligible broadening of the energy width of the spin waves. Blue dashed lines are fits with fixed $\gamma = 2\Gamma = 0.11$ meV and red lines are fits with optimized parameter $\gamma$. The $\chi^2$ of the blue lines with negligible damping is consistently larger than that of red lines. Blue dashed lines are also less accurate in describing the observed spectral features of the data, such as the broad high-energy tail of the spin wave (panels (c),(f)). Figure~\ref{supp3} shows similar comparison of the same 1D data cuts with the corresponding cuts from the 2D fits with variable (red soid lines) and fixed (blue dashed lines). A visible improvement of the fit quality in both Fig.~\ref{supp6} and Fig.~\ref{supp3} corroborates that the non-negligible intrinsic damping $\gamma$ indeed exists at 5.5~K.

\end{widetext}

\end{document}